\documentclass[referee,pdflatex,sn-mathphys-num]{sn-jnl}% Math and Physical Sciences Numbered Reference Style

\usepackage{graphicx}%
\usepackage{multirow}%
\usepackage{amsmath,amssymb,amsfonts}%
\usepackage{amsthm}%
\usepackage{mathrsfs}%
\usepackage[title]{appendix}%
\usepackage{xcolor}%
\usepackage{textcomp}%
\usepackage{manyfoot}%
\usepackage{booktabs}%
\usepackage{algorithm}%
\usepackage{algorithmicx}%
\usepackage{algpseudocode}%
\usepackage{listings}%
\begin{document}

\title[Article Title]{Multiscale transitional flow in anisotropic nanoparticle suspensions revealed by 
time-resolved x-ray scatter microscopy}

\author[1]{\fnm{Kesavan} \sur{Sekar}}

\author[1]{\fnm{Viney} \sur{Ghai}}

\author[1,2,a]{\fnm{Reza} \sur{Ghanbari}}

\author[1,3]{\fnm{Marko} \sur{Bek}}

\author[4,5,6]{\fnm{Marianne} \sur{Liebi}}

\author[4,7]{\fnm{Aleksandar} \sur{Matic}}

\author[2,3]{\fnm{Ann~E.} \sur{Terry}}

\author*[2,3]{\fnm{Kim} \sur{Nyg{\aa}rd}}\email{kim.nygard@maxiv.lu.se}

\author*[1,2,3,7]{\fnm{Roland} \sur{K{\'a}d{\'a}r}}\email{roland.kadar@chalmers.se}

\affil[1]{\orgdiv{Department of Industrial and Materials Science}, 
\orgname{Chalmers University of Technology}, 
\postcode{SE-412 96}, \city{Gothenburg}, \country{Sweden}}

\affil[2]{\orgdiv{MAX IV Laboratory}, 
\orgname{Lund University}, 
\postcode{SE-224 84}, \city{Lund}, \country{Sweden}}

\affil[3]{\orgname{LINXS Institute of Advanced Neutron and X-ray Science},

\postcode{SE-224 84}, \city{Lund}, \country{Sweden}}

\affil[4]{\orgdiv{Department of Physics}, 
\orgname{Chalmers University of Technology}, 

\postcode{SE-412 96}, \city{Gothenburg}, \country{Sweden}}

\affil[5]{\orgdiv{Division of Photon Science}, 
\orgname{Paul Scherrer Institute}, 

\postcode{CH-5232}, \city{Villigen}, \country{Switzerland}}

\affil[6]{\orgdiv{Institute of Materials}, 
\orgname{Ecole Polytechnique F{\'e}d{\'e}rale de Lausanne (EPFL)}, 
\postcode{CH-1015}, \city{Lausanne}, \country{Switzerland}}

\affil[7]{\orgdiv{Wallenberg Wood Science Center (WWSC)}, 
\orgname{Chalmers University of Technology}, 
\postcode{SE-412 96}, \city{Gothenburg}, \country{Sweden}}

\affil[a]{Present address: NKT Technology Consulting, V{\"a}ster{\aa}s, Sweden}

\abstract{
Complex fluids transition from laminar to transitory flow above a critical control parameter, akin to 
their Newtonian counterparts. In a continuum mechanics sense, fluid elements follow the ensuing complex 
trajectories,  giving rise to secondary flows in terms of macroscopic vortices and patterns thereof. 
However, if we replace idealized fluid elements with actual anisotropic nanoparticles, 
would their trajectories still reveal the same spatiotemporal behavior as the macroscopic flow field?
This question is fundamental for complex fluids, where fully developed turbulence is suppressed by high 
viscosities and where understanding particle–flow coupling is central to transport, processing, and 
structure formation. To address this question, we develop small-angle x-ray scatter microscopy of 
unprecedented temporal resolution combined with polarized light imaging, thereby bridging seven orders 
of magnitude in lengtscales. Proof-of-principle is demonstrated on a classical stability problem, 
Taylor-Couette flow, of platelet-like graphene oxide nanoparticle and rod-like cellulose nanocrystal 
suspensions. The analysis shows hitherto hidden, markedly different multiscale dynamics underlying flow 
stability; while the platelet-like particles follow the wavy motion of the macroscopic, secondary 
flow field, the rod-like particles exhibit high-frequency motion that is uncorrelated with the vortex 
instabilities.
}

\keywords{Laminar-turbulent transition, Taylor-Couette instabilities, Multiscale spectral dynamics, 
Time-resolved small-angle x-ray scatter microscopy}

\maketitle

The transition from laminar to turbulent flow in complex fluids is a longstanding multiscale 
problem within classical physics, characterized by strong coupling between lengthscales several 
orders of magnitude apart~\cite{sreenivasan99,warhaft02}. One of the benchmark laboratory-scale experiments 
for characterizing such transitions is Taylor-Couette flow between concentric cylinders~\cite{taylor23}, 
with the inner cylinder rotating and the outer cylinder at rest. It can  be argued that pioneering studies 
date back to Maxwell's first unsuccessful attempts at flow-induced birefringence in a concentric cylinder 
setup, without any surviving details of his experimental setup whereby 'a solid cylinder could be 
made to rotate' inside a 'cylindrical box with a glass bottom'~\cite{Maxwell1873}, decades before the seminal 
experiments of Couette, Mallock, and Taylor~\cite{Donnely1991}. With increasing Reynolds number $Re$,  
describing the ratio between inertial to viscous forces, the fluid transitions from laminar to featureless 
turbulent flow via a cascade of increasingly complex, macroscopic vortex 
instabilities ~\cite{Wendt1933,gollub75,Andereck86,Muller97,Wereley1998}. In laminar Couette 
flow (LCF) a fluid element (in continuum mechanics sense) follows a simple circular trajectory. 
In the presence of flow disturbances at a critical $Re$, the trajectories will re-organize at 
first into three-dimensional helicoidal trajectories forming counter-rotating toroidal vortices, known 
as Taylor vortex flow (TVF), separated by radial outflow and inflow regions~\cite{Wereley1998}. Further 
increasing $Re$ results in wavy vortex flow (WVF), modulations thereof (modulated wavy vortices, MWV), 
chaotic merging and splitting of vortices (chaotic wavy vortices, CWV), before the onset of turbulent 
flow regimes in the form of turbulent wavy vortices (TWV) and turbulent Taylor vortices (TTV). We note 
that this simple pattern-development scenario is characteristic for both Newtonian fluids and the complex 
ones studied here, although (de)stabilization of non-Newtonian fluids can induce new instability 
modes~\cite{ghanbari24}. 

Considering the particular case of anistropic nanoparticles in the aforementioned complex transitional flow 
fields, several aspects must be considered. Generally, a mechanistic understanding of 
their non-equilibrium flow behavior requires multiscale structural 
characterization~\cite{vermant05,kadar21}. 
Beyond the onset of instabilities, the presence of spatiotemporally varying, macroscopic 
flow fields are expected to cause intricate, position-dependent alignment of dispersed anisotropic 
particles~\cite{assen22}. Depending on nanoparticle morphology and flow field, different nanoparticle 
symmetry axes may align with the flow as described below. However, while the macroscopic 
flow field can be resolved with various flow visualization or particle image velocimetry methods, these 
observations take place significantly above lengthscales that can be associated with the smallest vortical 
structures in transitional flows~\cite{tokgoz12}.

Spatially resolved small-angle x-ray scattering (hereafter denoted SAXS microscopy) has developed into 
an important experimental technique for mapping of nanoscale 
constituents~\cite{bunk09,feldkamp09,liebi15,schaff15}, widely applicable 
to different aggregation states of the sample and complex in-situ experiments. 
Indeed, Philippe et al.~\cite{philippe12} reported pioneering SAXS microscopy experiments on suspensions 
of platelet-like clay nanoparticles in Taylor-Couette flow, thereby identifying time-averaged alignment 
of the nanoparticles' director and their effect on the onset of instabilities. However, the modest 
temporal resolution in state-of-the-art SAXS microscopy of about a minute has so far limited the 
technique to spatiotemporal mapping of relatively slow processes, such as assembly of nanoparticle 
superlattices~\cite{yu18} or in-operando electrode sodiation~\cite{olsson25}.

Here we develop SAXS microscopy of unprecedented, millisecond temporal resolution, making use of the extreme 
x-ray flux offered by diffraction-limited synchrotron sources~\cite{eriksson14} and analysis in the 
frequency domain. Inspired by recent advances in multiscale analysis of complex fluids under shear 
flow~\cite{dunderdale20,bianco24,ghanbari24b}, we further quantify the macroscopic Taylor-Couette flow 
patterns 
and their dynamics by polarized light imaging (PLI), utilizing intrinsic flow-induced birefringence of 
the nanoparticle suspensions~\cite{ghanbari24}. We can thereby address an outstanding multiscale challenge 
in transitional Taylor-Couette flow: if we conceptually replace a fluid element (in the continuum sense) 
with an anisotropic nanoparticle, will we find the same spatiotemporal motion at macroscopic and 
nanoscopic scales? By combining PLI and SAXS microscopy observations of two nanostructured fluids with 
platelet- and rod-like nanoparticle morphologies in Taylor-Couette flow, we show for the first time 
instability modes where transitional flow effects at nanoparticle level co-exist with molecular motion 
effects, apparently depending on the particle morphology and/or aspect ratio. 

\subsection*{Custom Taylor-Couette-PLI-SAXS setup}
\begin{figure}[t]
\centering
\includegraphics[width=\textwidth]{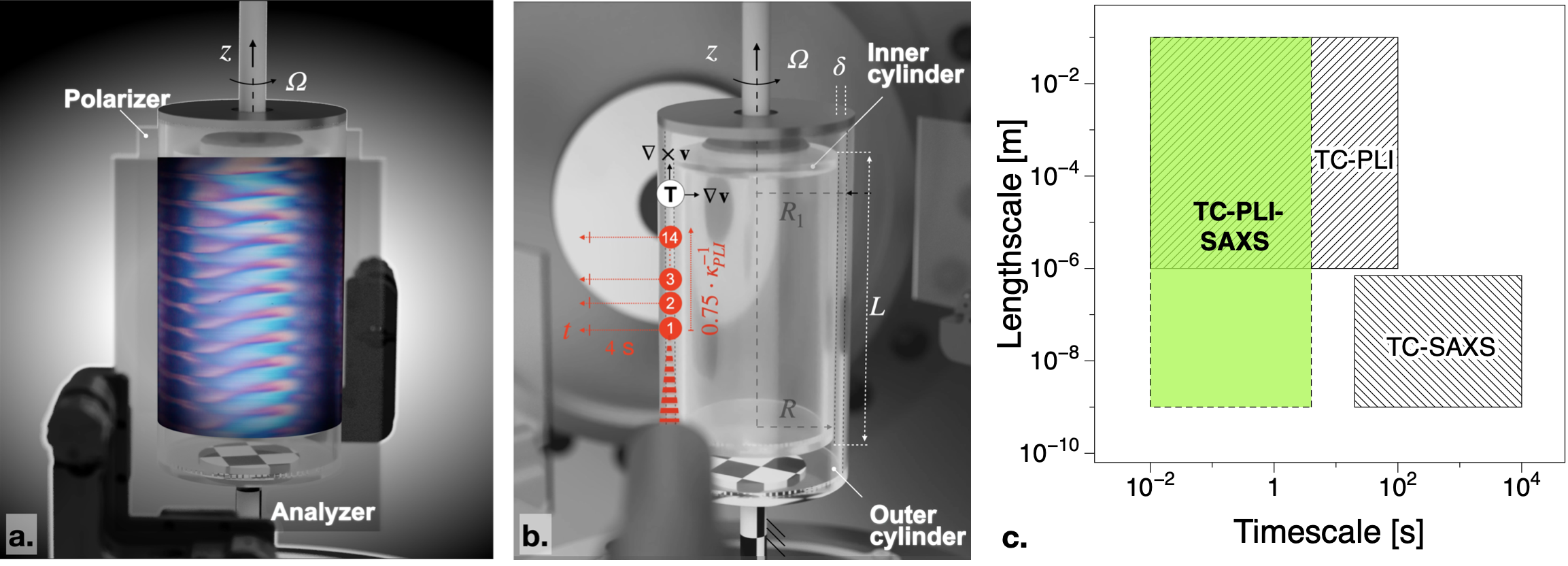}
\caption{Custom Taylor-Couette experimental setup for multiscale analysis of nanoparticle suspensions 
in supercritical flows. 
\textbf{a}, Illustration of the PLI visualization setup consisting of transparent concentric cylinders 
and cross-polarizer setup illuminated from the background with LED lights (not shown) and as viewed 
from the optical visualization camera (not shown), exemplifying wavy vortex flow. 
\textbf{b}, Tangential configuration (denoted by a circled T) of the SAXS experiment on the same flow cell,  
during which a set of fourteen SAXS time series are collected as function of position along the vorticity 
direction. 
\textbf{b}, Accessible length and time scales of the present PLI-SAXS study (green), compared to previous 
state-of-the-art PLI~\cite{ghanbari24} and SAXS~\cite{philippe12} experiments in Taylor-Couette flow. 
The velocity gradient and vorticity directions are denoted by $\nabla \mathbf{v}$ (or 2) 
and  $\nabla \times \mathbf{v}$ (or 3), respectively, where $\mathbf{v}$ is the velocity vector (direction 
denoted by 1), positioned orthogonally to the planes formed by the afore mentioned axes.
}
\label{FIG:setup}
\end{figure}

The current experiment is based on a custom Taylor-Couette flow cell presented in 
Figs.~\ref{FIG:setup}\textbf{a} and \ref{FIG:setup}\textbf{b}, where throughout this study we make use 
of rotating inner and stationary outer cylinders. In this geometry, PLI and SAXS experiments are carried 
out in different configurations~\cite{vermant05,eberle12}. On the one hand, PLI is employed in the 
radial configuration, visualizing the macroscopic flow in the plane defined by the velocity 
($\mathbf{v}$ or 1) and vorticity (\textit{z},  $\nabla\times\mathbf{v}$, or 3) directions. 
On the other hand, SAXS data are collected in the tangential configuration, probing the scattering plane 
defined by the velocity gradient ($\nabla\mathbf{v}$ or 2) and vorticity directions, while averaging the 
data over the velocity direction. We note that the choice of SAXS configuration is usually based on 
nanoparticle morphology; platelet-like particles predominantly align normal to the velocity gradient and 
are thus optimally studied in tangential configuration~\cite{philippe12,poulin16}, whereas rod-like 
particles align in the velocity direction and are hence preferably studied in radial scattering 
configuration~\cite{pignon21,munier22}. Nevertheless, in the presence of secondary flows as studied here, 
the tangential scattering configuration is suitable for both particle morphologies  
(see Supplementary Fig.~\ref{FIG:checks} for a quantitative comparison of scattering configurations).   
We emphasize that combined PLI and SAXS microscopy in Taylor-Couette geometry on a common millisecond 
timescale, as developed here and depicted in Fig.~\ref{FIG:setup}\textbf{c}, is critical to address 
transitional flow across lengthscales. 

We study two morphologically distinct nanoparticle suspensions - platelet-like graphene oxide (GO) and 
rod-like cellulose nanocrystals (CNC). Their static nanoscale structure can be inferred from time-averaged 
SAXS data collected under laminar Couette flow (LCF) as presented in Supplementary Fig.~\ref{FIG:SAXS_LCF};  
whereas the GO data decrease monotonously with scattering vector modulus $q$, without any signature 
of liquid-crystalline (or even short-range) ordering observed for GO at higher concentrations 
either under flow~\cite{poulin16} or in equilibrium~\cite{xu11}, indicating that we are probing individual 
nanoparticles, the CNC data exhibit distinct maxima in the structure factor at 
$q_\mathrm{max} \approx 0.16$~nm$^{-1}$, corresponding to local ordering at 
$2\pi/q_\mathrm{max} \approx 40$~nm interparticle distance. Throughout this study we average the 
SAXS data for both suspensions in the scattering vector range $q_0 = 0.13-0.20$~nm$^{-1}$. 

The prominent peak in the structure factor of the CNC SAXS data raises the question of probed lengthscale. 
For example, equilibrium suspensions exhibit slowing down of diffusivity at intermediate $q$, while 
retaining single-particle dynamics at larger $q$~\cite{nagele96,nygard16}, an effect sometimes colloquially 
coined de~Gennes narrowing following Ref.~\cite{degennes59} (despite the different underlying physics 
between diffusion of suspended nanoparticles and dynamics of atomic/molecular systems~\cite{ackerson82}). 
In the present case, however, the qualitatively different nanoparticle dynamics observed 
for platelet-like GO versus rod-like CNC (discussed below) are not due to local ordering, as 
evidenced by the $q$- and concentration-dependent CNC data of Supplementary Fig.~\ref{FIG:checks}.

\subsection*{Multiscale analysis of supercritical flow patterns}
\begin{figure}[t]
\centering
\includegraphics[width=\textwidth]{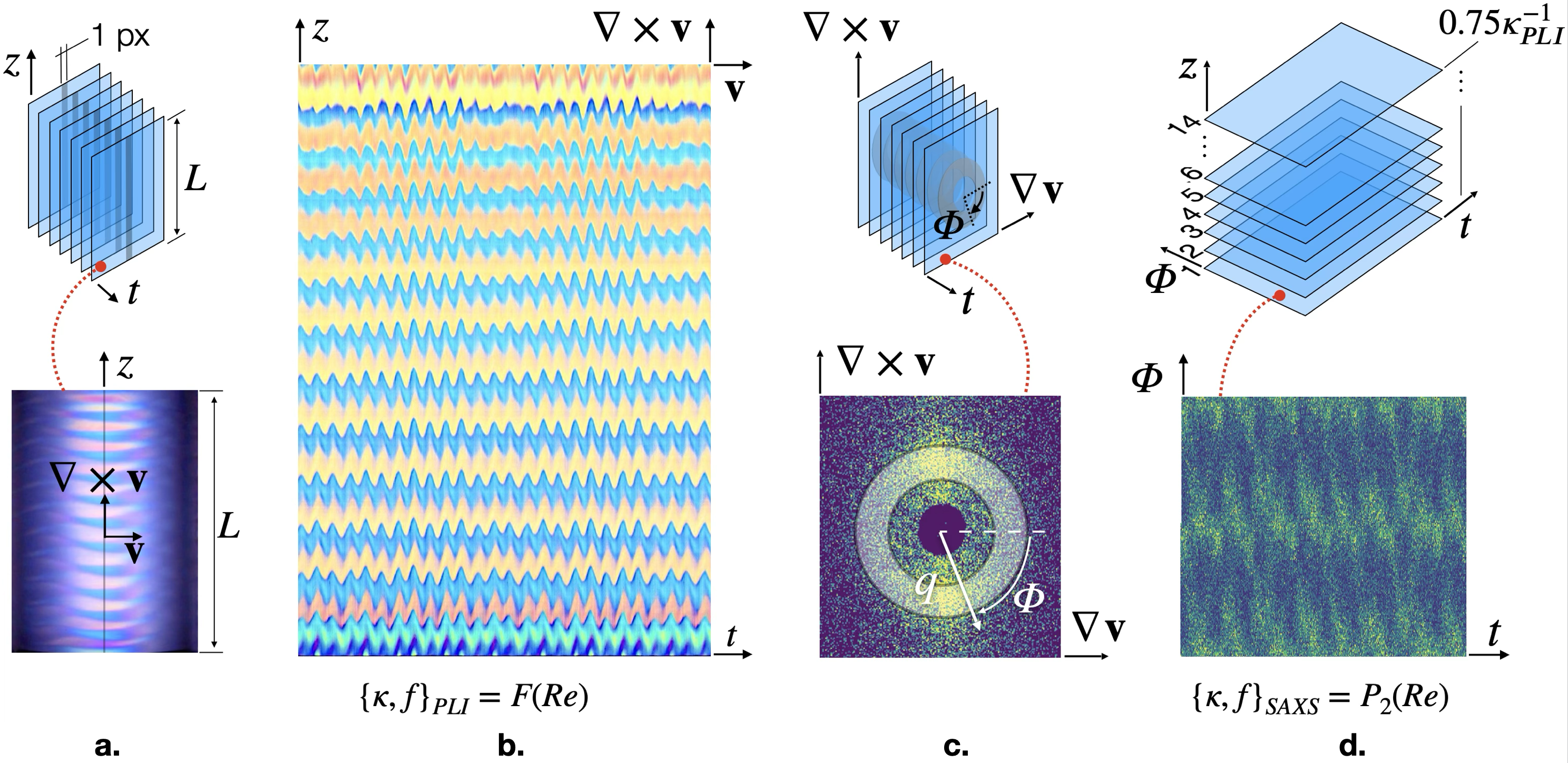}
\caption{Summary of multiscale data analysis of TC-PLI-SAXS experiments. 
\textbf{a}, PLI video recordings 
processed as a sequence of frames wherefrom a line of 1 pixel is extracted along the vertical ($z$) axis of 
the cylinders. 
\textbf{b}, Example of PLI space-time diagram constructed from the individual frames for modulated wavy 
vortices (MWV) in a CNC suspension. Following two-dimensional Fourier analysis, the macroscale characteristic 
wavenumber (spatial periodicity) and frequency(ies) (temporal periodicity), $\{\kappa, f\}$, of the MWV can 
be determined. 
\textbf{c}, SAXS data processing, starting from azimuthal integration of time series of scattering patterns over a 
range of scattering vector moduli $q$. 
\textbf{d}, From the azimuthally integrated data, the scattered intensity 
$I(\it\Phi,z,t)_{q_0}$ can be determined for each $z=ct.$ scan. Thereafter, for each time step the scattering 
anisotropy was determined using the Hermans orientation parameter $P_2$, with Fourier analysis thereof yielding 
nanoscopic, characteristic spatial and temporal frequencies.
}
\label{FIG:analysis}
\end{figure}

To make macroscale PLI flow pattern and SAXS nanoscale observations directly comparable, we have developed a 
common spatiotemporal analysis framework as illustrated in Fig.~\ref{FIG:analysis}. For PLI, single pixel 
lines along the $z$ axis of the cylinders are combined from the frames of video recordings 
(panel~\textbf{a}) to form macroscopic space-time diagrams (panel~\textbf{b}), exemplified for 
the CNC suspension subjected to MWV. This particular instability mode exhibits (i)~spatial periodicity 
along the vorticity direction as well as (ii)~high-frequency periodicity and (iii)~low-frequency modulation 
in the time domain. Spatial and temporal two-dimensional Fourier analysis thereof following 
Refs.~\cite{ghanbari24,dutcher09} yield the characteristic spatial wavenumber $\kappa$ and temporal 
frequencies $f_n$ versus Reynolds number $Re$. For SAXS microscopy, in turn, two-dimensional scattering 
patterns acquired at the same frame rate with PLI are azimuthally integrated over $q$ (panel~\textbf{c}) 
to construct azimuthal angle ($\it\Phi$) - time diagrams corresponding to each of the $z=ct.$ scans 
(panel~\textbf{d}). Expanding the standard approach for characterizing nanoparticle flows by small-angle 
scattering, see e.g. Refs.~\cite{eberle12,hakansson14,ghanbari24b}, we base our analysis on a {\em spatially} 
and {\em temporally} resolved variant of the Hermans orientation parameter~\cite{hermans39}, 
\begin{equation}
P_2(z,t) = \frac{\int_0^\pi \frac{1}{2}\left(3\cos^2\varPhi -1\right) I(\it\Phi,z,t)_{q_0} 
\sin\it\Phi d\it\Phi} {\int_0^\pi I(\it\Phi,z,t)_{q_0} \sin\it\Phi d\it\Phi}.
\label{Eq:P2}
\end{equation}
Here, $I(\it\Phi,z,t)_{q_0}$ denotes the spatially and temporally resolved annular scattering pattern along the 
azimuthal angle $\it\Phi$ (defined clockwise with respect to the velocity gradient direction, see 
Fig.~\ref{FIG:analysis}\textbf{c}), averaged over the above specified scattering vector range $q_0$. We 
note that $P_2(z,t)$ provides an instantaneous, ensemble-averaged, and local measure of the preferred 
alignment and degree thereof of the suspended nanoparticles. The current choice of azimuthal integration 
limits is particularly useful for studying Taylor-Couette instabilities; positive and negative values of 
$P_2(z,t) $ indicate that the projection of the nanoparticles' director in the scattering plane 
preferentially aligns in the vorticity and velocity gradient directions, respectively. 
Alternatively we could determine the preferential alignment of the nanoparticles' 
director, $\varphi(z,t)$, as outlined in the Methods section and shown in Supplementary Fig.~\ref{FIG:checks}. 
In either case, subsequent Fourier analysis allows quantitative comparison with macroscopic PLI data. 
The SAXS scans in the $z$ direction are scaled to the wavenumbers 
determined from PLI analysis, to cover one and a half  adjacent vortices ($\Delta z$ is a pair 
of vortices).

\subsection*{Onset of instabilities}
\begin{figure}[t]
\centering
\includegraphics[width=\textwidth]{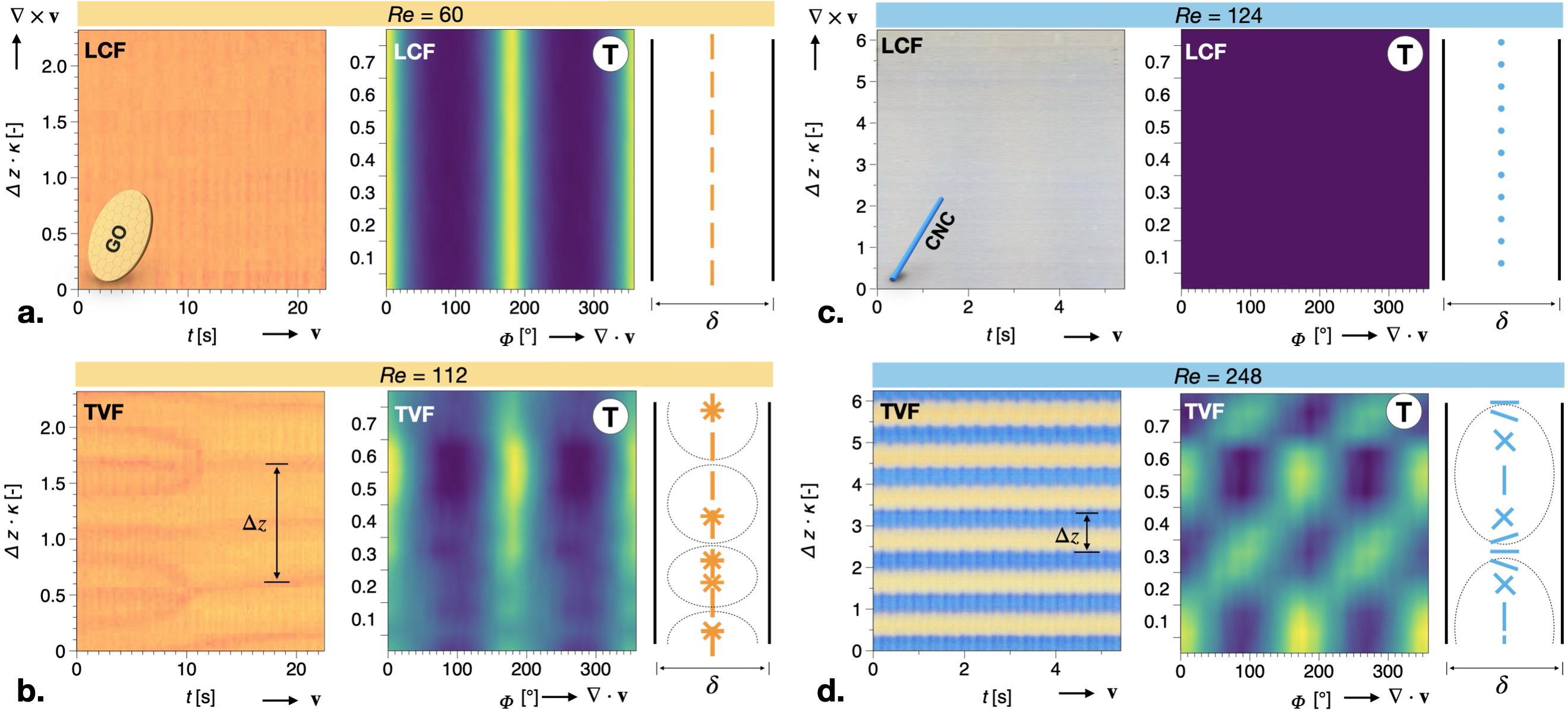}
\caption{Multiscale observation of flow transitions. The data present the transition from 
(\textbf{a},\textbf{c})~laminar Couette (LCF) to (\textbf{b},\textbf{d})~Taylor vortex 
flow (TVF) for (\textbf{a},\textbf{b}) the GO and (\textbf{c},\textbf{d}) the CNC suspension. 
Each figure compares PLI visualization (left), temporally averaged azimuthally integrated SAXS data 
$\langle I(\it\Phi,z)_{q_0}\rangle_t$ (middle), and a schematic picture of nanoparticle alignment based 
on the above data (right).
Data collected in tangential configuration are highlighted with a circled T.  
}
\label{FIG:experiment}
\end{figure}

As alluded to in the introduction, transitional flow proceeds via a sequence of increasingly complex 
macroscopic vortex instabilities with increasing Reynolds number $Re$. For rheologically complex fluids 
as studied here (see Supplementary Fig.~\ref{FIG:viscosity} for shear viscosities of the investigated 
suspensions), the detailed onset and order of the instability transitions depend on an intricate 
interplay between shear-thinning and elasticity effects, a detailed discussion of which is beyond the 
scope of the present paper. Nevertheless, we summarize a few aspects of their flow stability. Laminar 
Couette flow (LCF) transitions at a critical Reynolds number $Re_{cr1}$ to Taylor vortex flow (TVF), 
a time-invariant axisymmetric instability mode consisting of counter-rotating toroidal vortices 
defined by one characteristic wavenumber $\kappa$ along $z$ but no temporal frequency $f_n$. 
Fig.~\ref{FIG:experiment} exemplifies the LCF-TVF transition as visualized by PLI and SAXS. The presence 
of GO has a destabilizing effect on the onset of instabilities, with 
$Re_{cr1}^{GO}<Re_{cr1}^{Newt} \approx 144$. 
Furthermore, the transition to TVF is initially an elasticity-modified pattern as mentioned above. 
Supercritical modes are Newtonian-like, although significant differences are observed for other 
boundary conditions~\cite{sekar25}. CNC, in turn, has a stabilizing effect on the flow, with 
$Re_{cr1}^{CNC}>Re_{cr1}^{Newt}$, while shear-thinning effects lead to smaller characteristic TVF 
wavenumbers than for the reference Newtonian case. All supercritical flow patterns are Newtonian-like, 
albeit with onset and range of $Re$ differing considerably compared to the Newtonian case ~\cite{ghanbari24}. 

Before proceeding further, we briefly discuss the onset of the present instabilities and validate 
our observations with expected results based on the available scientific literature. For GO and CNC 
subjected to LCF, the space-time diagrams in the left panel of Figs.~\ref{FIG:experiment}\textbf{a} and 
\ref{FIG:experiment}\textbf{c} (radial configuration) confirm the absence of any 
discernible macroscopic flow patterns. Focusing instead on the time-averaged azimuthal SAXS data of the 
middle panel, GO shows pronounced peaks at ${\it\Phi} =0$ and $180^\circ$, while for CNC no significant 
scattering is observed in the tangential configuration. Qualitatively speaking, the 
nanoparticles primarily scatter x-rays parallel to their short axis, providing direct access to their 
alignment. This confirms the expected scenario; GO and CNC nanoparticles predominantly align with their long 
axes perpendicular to $\nabla\textbf{v}$ and parallel to $\mathbf{v}$, respectively. Turning next to GO 
subjected to Taylor vortex flow, the nanoplatelets preferentially align vertically (vorticity direction, 
$\nabla \times \mathbf{v}$) in the radial inflow / outflow regions, akin to previous observations for 
suspensions of platelet-like particles in TVF~\cite{philippe12}. The inside of the vortices can be 
identified by variations in the orientation direction, Fig.~\ref{FIG:experiment}\textbf{b}. We further 
note that the TVF pattern debuts with higher wavenumbers (decreased height of the vortices) and then 
settles for lower wavenumbers. The CNC data of Fig.~\ref{FIG:experiment}\textbf{d}, in contrast, shows 
distinct patterns in both PLI and SAXS visualization at the onset of TVF, with the nanoparticle director 
following the macroscopic vortex in its toroidal trajectory.

\subsection*{Supercritical instability modes}
\begin{figure}[t]
\centering
\includegraphics[width=\textwidth]{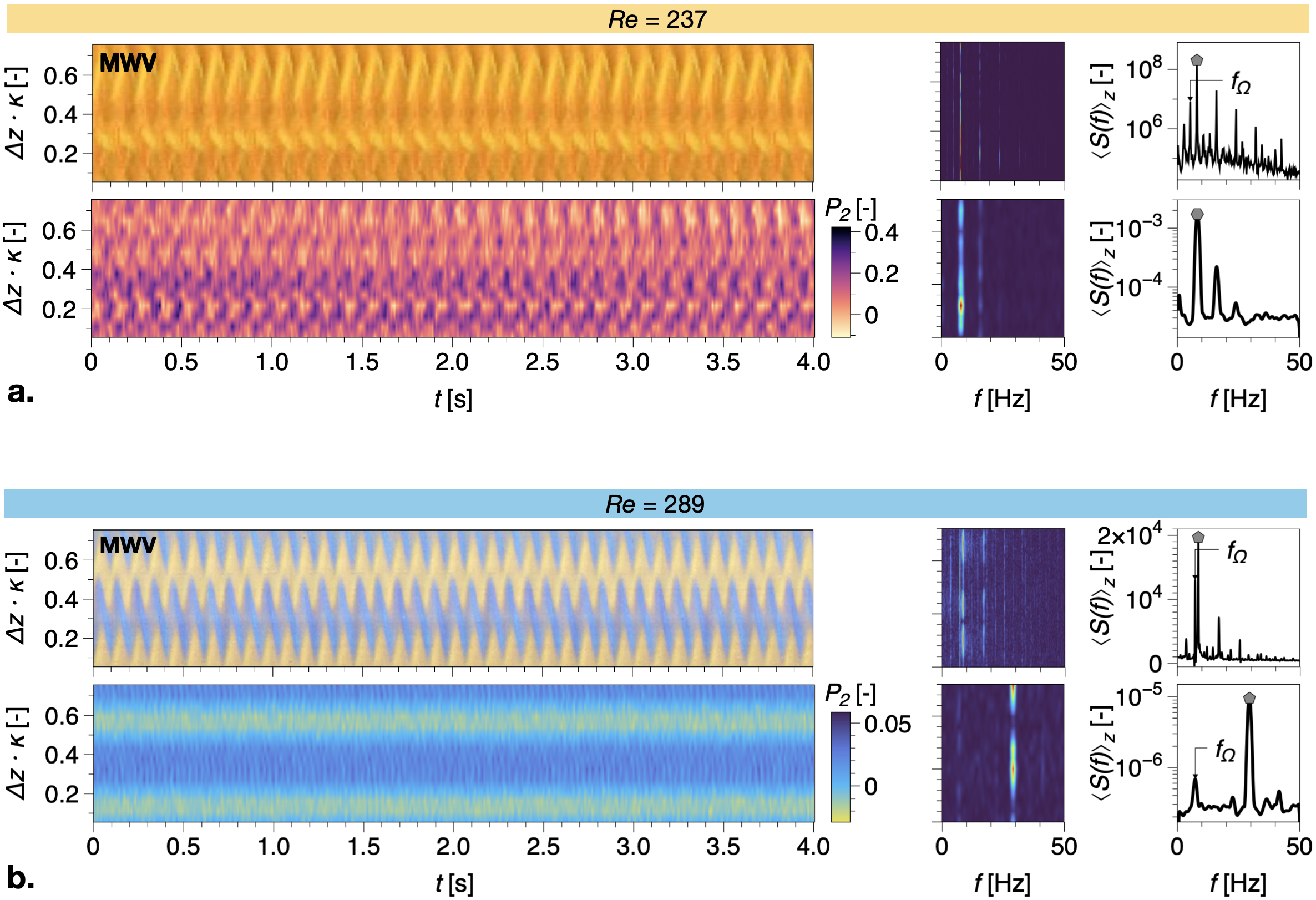}
\caption{Multiscale spatiotemporal analysis of transitional flow in anisotropic nanoparticle 
suspensions. Data are presented for (\textbf{a})~GO and (\textbf{b})~CNC suspensions subjected to 
modulated wavy vortex flow (MWV). 
The left panel presents the macroscopic space-time plot obtained by PLI (top) and 
the spatiotemporally resolved nanoscopic Hermans orientation parameter $P_2(z,t)$ determined by SAXS 
(bottom). The middle and right panels 
display spatially resolved $S(z,f)$ and averaged $\langle S(f)\rangle_z$ power spectral densities at 
both macroscopic (top) and nanoscopic 
lengthscales (bottom), determined by Fourier analysis from the data of the left panel. The rotational 
frequency $f_\Omega$ of the Taylor-Couette 
cell and the characteristic frequency $f_n$ (pentagon) are also depicted in the right panel. Note the 
common spatial scale for both PLI and SAXS. 
}
\label{FIG:SAXS}
\end{figure}

Next we proceed to the sequence of vortex instabilities, as outlined briefly in the introduction and 
detailed in Table~\ref{TAB:instabilities}. Fig.~\ref{FIG:SAXS} compares modulated wavy vortex flow at 
similar $Re$ in both suspensions. In the left panel of Fig.~\ref{FIG:SAXS}\textbf{a} we apply the above 
developed multiscale characterization on the platelet-like GO suspension, noting a remarkable similarity 
between the macroscopic flow field and the alignment of the nanoparticle director in terms of both spatial 
and temporal periodicity. For comparison we also present corresponding data for the rod-like CNC suspension 
in Fig.~\ref{FIG:SAXS}\textbf{b}, displaying markedly different spatiotemporal evolution at macroscopic 
and nanoscopic scales; while the CNC director exhibits the same spatial periodicity as the macroscopic flow 
field, it shows no signature of large-scale wavy dynamics. Further examples underscoring the discrepancy 
are presented in Supplementary Fig.~\ref{FIG:multiscale_complementary}. 

The spatially and temporally resolved nanoscale orientation parameters above are determined from sequences 
of position-dependent SAXS time series. While data synchronization by cross correlation is possible 
(at least for GO), we have concluded that the similarity between PLI and SAXS is captured without the 
need for synchronization of the relatively high frequency of the wavy oscillations.  Nevertheless, 
in order to facilitate quantitative analysis of the multiscale dynamics, we present in the middle panels 
of Fig.~\ref{FIG:SAXS} the power spectral density of the spatiotemporal orientation parameter, 
\begin{equation}
S(z,f) = \lim\limits_{t_0 \to \infty} t_0^{-1}\vert \mathcal{F} [P_2(z,t) - \langle P_2(z)\rangle_t] \vert^2, 
\label{eq1}
\end{equation}
where $f$ denotes the frequency and $\mathcal{F}$ the temporal Fourier transform taken over the time 
window $t_0$. Here we subtract for convenience the temporally averaged orientation parameter, 
$\langle P_2(z) \rangle_t = t_0^{-1} \int_0^{t_0}P_2(z,t) dt$, in order to suppress the zeroth 
Fourier component. While such Fourier analysis is readily applicable for PLI as it requires 
acquisition frame rates that are accessible in many commercial optical cameras, the temporal resolution 
has been so far a limiting factor in SAXS analysis. For both GO and CNC we observe distinct spectral 
dynamics, markedly different from the broadband power spectral density $\propto f^{-2}$ expected for 
(rotational) Brownian motion. Since we observe no shift in the characteristic frequencies with position 
along the vorticity direction, we also present in the right panels the spatially averaged power spectral 
densities $\langle S(f)\rangle_z = (\kappa/0.75)\int_0^{0.75\kappa^{-1}}S(z,f)dz$.  
For GO we observe a single $f_n \approx 8$~Hz excitation and its higher harmonics, whereas for CNC we 
observe an $f_n \approx 29$~Hz excitation and a weak signature of the $f_{\Omega} \approx 7$~Hz rotation 
of the inner cylinder of the Taylor-Couette cell. In both cases, the nanoparticle excitations constitute 
motion centered around the mean director schematically depicted in the right panel of 
Figs.~\ref{FIG:experiment}\textbf{b} and \ref{FIG:experiment}\textbf{d}, distinctly different from the 
tumbling, precessing, and rolling motion expected for such suspension in shear flow between planar 
surfaces~\cite{huang12}. 

\begin{figure}[t]
\centering
\includegraphics[width=0.6\textwidth]{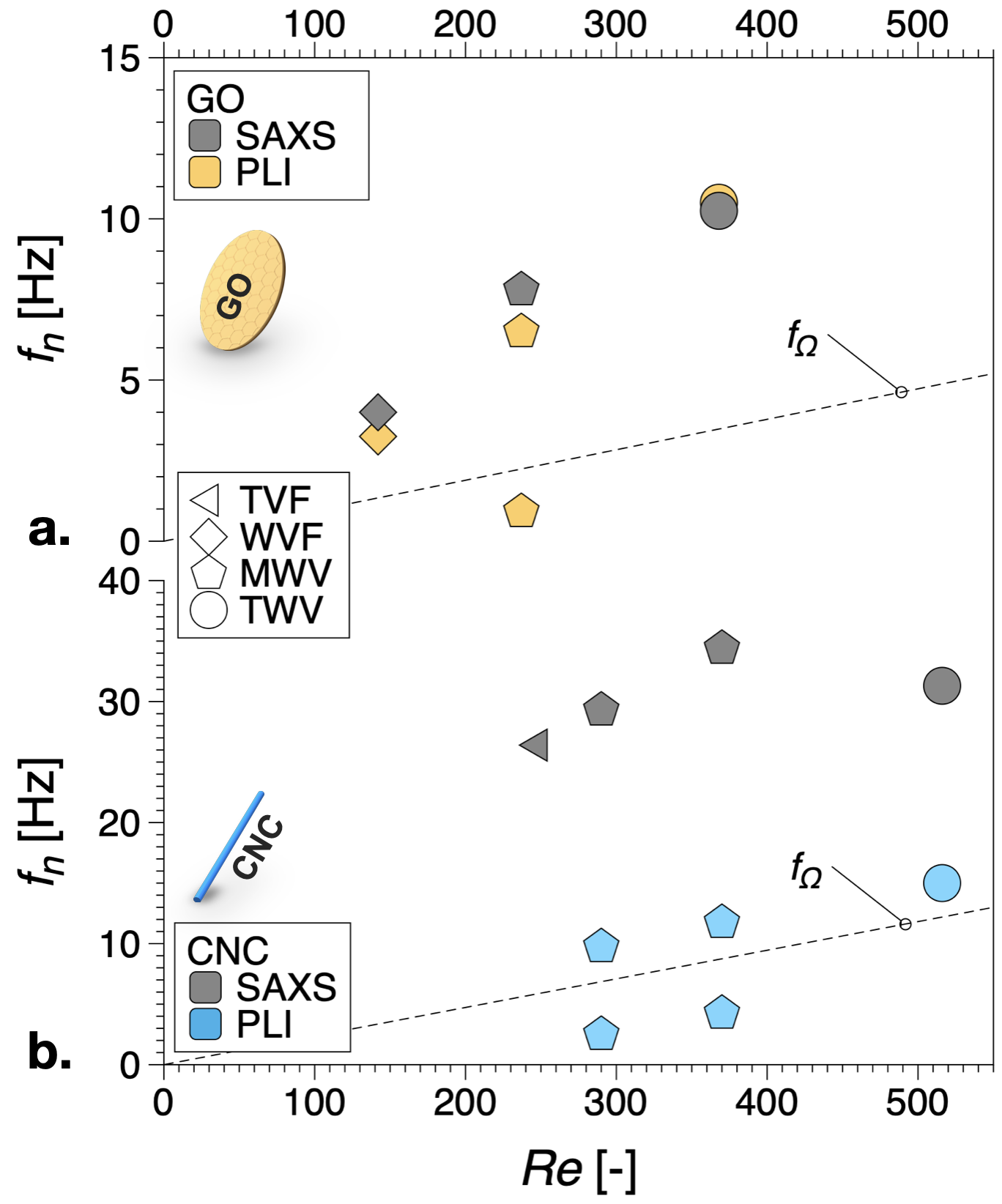}
\caption{Multiscale dynamics of nanoparticle suspensions subjected to transitional flow. 
Both nanoscopic (SAXS) and macroscopic (PLI) data are presented for (\textbf{a})~GO and (\textbf{b})~CNC 
as the characteristic frequencies 
$f_n$ versus Reynolds number $Re$. The dashed line depicts the rotational frequency $f_{\Omega}$ 
of the Taylor-Couette cell. Note the different scale on the frequency axes for the GO and CNC data.}
\label{FIG:freq}
\end{figure}

To further emphasize this remarkable finding, we correlate in Fig.~\ref{FIG:freq} the characteristic 
macroscopic and nanoscopic frequencies as probed by PLI and SAXS, respectively. The multiscale 
dynamics of GO and CNC are found to behave markedly different across all instability 
modes ivestigated. The GO nanoplatelets follow the wavy motion of the macroscopic 
flow field. For TVF we cannot identify a single characteristic temporal frequency, as expected based 
on PLI data, whereas the slower modulation frequency of the MWV instability seems to fall outside the 
spatiotemporal window of the present SAXS experiment. In contrast, the CNC director oscillates with a 
significantly higher frequency, apparently neither correlated with the macroscopic flow field nor 
a higher harmonic of the rotational frequency $f_{\Omega}$. We note that 
the CNC data are reminiscent of recent simulation results of rod-like particles aligning along the local 
tangent of the Taylor-Couette geometry~\cite{assen22}, although the latter results were obtained for 
different concentration and dimensions of the suspended particles. We foresee that these experimental 
results will stimulate further simulation studies on this topic.  

\section*{Discussion}
We can partly understand the strikingly different multiscale spatiotemporal behavior of 
platelet-like GO versus rod-like CNC supensions by considering the balance between flow-induced rotational 
motion, in the form of the macroscopic wavy mode acting as a 'forcing' frequency on the nanoparticles, 
and de-orientation due to Brownian motion, as expressed by the rotational diffusion coefficient $D_r$. 
Together they constitute a rotational supercritical flow Péclet number,  
$Pe_{r,sc}^{(w)} = f_{n,PLI}/D_r$. 
In this interpretation, $Pe_{r,sc}^{(w)} \gg 1$ implies that supercritical-flow-induced effects dominate and 
therefore the particles  align with the flow, in contrast to $Pe_{r,sc}^{(w)} \ll 1$ 
for which Brownian motion dominates. Interestingly, we obtain $Pe_{r,sc}^{(w)} \propto 10^1$ for GO and 
$ \propto 10^{-1}$ for CNC, respectively, clearly separating the two regimes.

The detailed impact of the elucidated multiscale dynamics on the flow stability and the classical 
macroscopic interpretation thereof in terms of elastic and shear thinning effects remains to be determined. 
The fact that GO shows evidence of fluid elasticity effects at the transition to TVF while mostly 
shear-thinning effects are plausibly assigned to CNC is intriguing. Likewise, despite both GO and CNC 
suspensions showing Newtonian-like flow patterns, GO stabilizes and CNC destabilizes the onset of 
instabilities, akin to the platelet-like Oldroyd-A and rod-like Oldroyd-B models~\cite{hillebrand25}. 
Understanding this multiscale interplay could have significant ramifications, e.g., 
for suppression or enhancement of certain modes using nanoscale material design.

Finally, vortex instabilities in Taylor-Couette flow also serve as a model for instability-induced 
pattern formation~\cite{cross93}, a phenomenon of orderly self-organization ranging from snow 
flakes~\cite{benjacob90} to cortical convolutions~\cite{tallinen16}. These are often multiscale in 
nature, with some degrees of freedom at small lengths participating in pattern formation at macroscopic 
scales, although explicit multiscale modeling of pattern formation is scarce~\cite{wurthner22}. A key 
question then is which small-scale degrees of freedom are relevant for macroscopic pattern formation 
and which can be coarse-grained out?  Given the markedly different multiscale dynamics of GO and CNC 
reported above, we foresee explicit studies on how nanoparticle morphology affect macroscopic pattern 
formation in transitional flow.

\section*{Methods}

\subsection*{Materials}
Graphene oxide (Highly acidic 20~\% aqueous paste from Abalonyx; short axis $d = 1 \pm 0.5$~nm; long axis 
$l = 2 \pm 1~\mu$m) was dispersed in Milli-Q water using an Elmasonic P30H bath sonicator at 37~KHz 
frequency for 30 minutes. Cellulose nanocrystals (CelluForce NCC\textsuperscript \textregistered NCV100-NASD90; 
$d = 6 \pm 2$~nm; $l = 230 \pm 70$~nm) were dispersed in Milli-Q water using Branson Sonifier 450 (20~kHz 
ultrasonic horn with a 1/2” diameter probe tip) operating at an amplitude of 60~\% with a duty cycle of 
60~\% for 30 minutes. Both suspensions were placed in an ice bath to prevent sample heating and 
replenished frequently. 

Supplementary Fig.~\ref{FIG:viscosity} presents the shear viscosity $\eta$ of both suspensions. 
The GO data are well described by a power-law model, $\eta (\dot{\gamma}) = K \cdot \dot\gamma^{n-1}$, 
where $\dot\gamma$ is the shear rate, $K$ a prefactor, and $n$ the power index. 
The CNC data, in turn, can be described by the Carreau-Yasuda model~\cite{Larson1999}, 
$\eta(\dot\gamma)= \eta_\infty + (\eta_0 - \eta_\infty) \left[ 1 + (\lambda \dot\gamma)^a\right]^\frac{n-1}{a}$,
with $\eta_0$ and $\eta_\infty$ denoting the viscosities at zero and infinite shear rates, respectively, 
$\lambda$ the characteristic time, $a$ the transition parameter, and $n$ the power index. 

Rotational diffusion coefficients were estimated~\cite{Larson1999} based on the classical thin-disk 
approximation in Stokes flow for GO, $D_r^{GO} = 3 k_B T/8 \pi \eta l^3$, 
and on the slender-rod approximation for CNC, $D_r^{CNC} = 3 k_B T(\ln(l/d))/\pi \eta l^3$, 
where $k_B = 1.38 \times 10^{-23}$ J/K and all experiments carried out at room temperature, $T = 296$ K.  
The ensuing rotational diffusion coefficients, $D_r^{GO} \approx 0.16$ and $D_r^{CNC} \approx 94$~s$^{-1}$, 
are in good agreement with previous experimental estimations~\cite{kim17,vanrie19}. 

\subsection*{Custom flow cell}
The experiments were performed using a custom Taylor-Couette cell specifically designed for flow 
visualization. The cell was mounted on an Anton Paar MCR702 Multidrive rheometer in dual motor 
configuration and counter-movement mode, with only the inner cylinder rotating.
In the combined PLI and SAXS experiment both cylinders were made of polycarbonate in order to minimize 
parasitic SAXS scattering, while the PLI data presented above were collected in a separate PLI experiment 
using a polycarbonate inner and glass outer cylinder for improved visibility. The inner and outer cylinders 
have radii of $R=20.5$ and $R_1=22.5$~mm, respectively, with the gap ratio $\epsilon = R/R_1 = 0.91$, 
while the cell height is $L=68.7$~mm 
(cf. Fig.~\ref{FIG:setup}\textbf{b} for geometrical definitions of the flow cell). We determined the 
dimensionless Reynolds number, $Re = \rho R\it\Omega\delta/\eta$, using rheological data, with $\delta$ 
denoting the gap width, $\it\Omega$ the angular velocity of the inner cylinder, and $\rho$ the density 
of the suspension. As an alternative quantification of inertial to viscous forces, we also present in 
Table~\ref{TAB:instabilities} the dimensionless Taylor number, $Ta = \rho^2 R\it\Omega^2\delta^3/\eta^2$.  

\subsection*{Polarized light imaging (PLI)}
The PLI experiment was carried out using an optical setup with two orthogonal linear polarizers placed 
upstream and downstream of the custom Taylor-Couette flow cell. In brief, we recorded flow visualization 
videos in the velocity-vorticity plane versus Reynolds number at 100~Hz for 1800~s, from which we generated 
space-time plots. We carried out spatial and temporal analysis by performing two-dimensional fast Fourier 
transforms (FFTs) of the space-time plots and averaging over the temporal and spatial frequencies, 
respectively. The use of intrinsic flow-induced birefringence of the nanoparticle suspensions yields 
similar characteristic frequencies as standard visualization particles~\cite{dutcher09}. For more details 
on the PLI experiments, we refer to Ref.~\cite{ghanbari24}. 

\subsection*{Small-angle x-ray scattering (SAXS)}
The SAXS experiment was carried out at the ForMAX beamline of the MAX IV Laboratory~\cite{nygard24}, 
making use of the high-flux multilayer monochromator setup and the EIGER2 X 4M~\cite{donath23}  
photon-counting pixel detector $\approx 1.8$~m downstream of the sample. We employed an incident x-ray 
beam of $\approx 5\times 10^{14}$~photons~s$^{-1}$, with an energy of 16.3~keV and a moderate beam size 
of $\approx 200\times 200~\mu$m$^2$ at the sample position. For each sample and Taylor-Couette instability, 
we mapped the instabilities along the vorticity direction in 14 steps in the range of 
$0-0.75\kappa_{PLI}^{-1}$, i.e., covering approximately one and a half vortices, with $\kappa_{PLI}$ 
being the wavenumber of each instability as determined from spatial Fourier analysis of the PLI data  
($\kappa_{PLI}^{-1} \approx 5-10$~mm depending on sample and instability mode).  
In each position, we sampled a SAXS time series at 100~Hz for 4~s, during which we observed 
radiation damage in neither SAXS nor torque data. Finally, we estimated the power spectral density of the 
spatially and temporally resolved orientation parameter $P_2(z,t)$ using Welch's method of overlapped 
segment averaging.

As an alternative to SAXS analysis based on the spatiotemporal Hermans orientation parameter $P_2(z,t)$ 
of Eq.~(\ref{Eq:P2}), we also determined the local and instantaneous angle of the nanoparticles’ director, 
$\varphi(z,t) = \arctan(\Im(a_1)/\Re(a_1))+\pi/2$, the degree thereof, $\vert a_1 \vert / \vert a_0 \vert$, 
and the scattering power, $\vert a_0 \vert$, where $\Re$ and $\Im$ denote real and imaginary parts of a 
complex quantity and $a_i = a_i(z,t)$ the ith complex FFT component of $I(\varPhi,z,t)_{q_0}$ along the 
azimuthal dimension $\varPhi$ (see Supplementary Fig.~\ref{FIG:checks}\textbf{a}). Such FFT-based 
analysis will be highly advantageous in future applications of the present methodology to 
two-dimensional scans composed of $\geq 10^5$ SAXS patterns. 

The Taylor-Couette geometry warrants a brief discussion. Both PLI and SAXS probes an 
orientation projection of a complex, three-dimensional and time-dependent flow field, averaged over 
the velocity gradient and velocity directions in the radial and tangential configurations, respectively. 
In the tangential case, SAXS analysis is further complicated by the curvature of the geometry 
($\leq 17^\circ$) along the x-ray beam path, partly smearing out the signal from nanoparticle alignment. 
Nevertheless, the identical nanoscale dynamics observed in both scattering configurations, as exemplified 
in Supplementary Fig.~\ref{FIG:checks}\textbf{b}, support the interpretation that the nanoparticle 
director dynamics probed by SAXS in tangential configuration can be quantitatively compared to the 
macroscopic dynamics probed by PLI in radial configuration.

\backmatter

\bmhead{Acknowledgements}
Research conducted at MAX IV, a Swedish national user facility, is supported by the Swedish Research 
Council under contract 2018-07152, the Swedish Governmental Agency for Innovation Systems under contract 
2018-04969, and Formas under contract 2019-02496. The work was supported by Chalmers Foundation through 
the project Chalmers Center for Advanced Neutron and X-ray scattering techniques in cooperation with MAX IV. 
K.S., V.G., M.B., and R.K. are grateful for the financial support of the Wallenberg Wood Science Center 
(WWSC), FibRe Vinnova Competence Center, 2D-TECH Vinnova Competence Centre and Chalmers Areas of Advance 
Materials Science, Nano and Production. The authors thank Alexandra Aulova, Ases A. Mishra, and Sajjad 
Pashazadeh for assistance during the SAXS experiments. The authors are grateful to Prof. Gareth McKinley 
for helpful remarks on dimensionless scaling.

%\bibliography{TC_transitional_flow}% common bib file

%%%%%%%%%%%%%%
%%% TABLES %%%
%%%%%%%%%%%%%%

\newpage

% TABLE 1
\begin{table}[t]
\centering
\caption{Main characteristics of the studied Taylor-Couette instabilities: laminar Couette flow (LCF), 
Taylor vortex flow (TVF), wavy vortex flow (WVF), modulated wavy vortex flow (MWV), and turbulent wavy 
vortex flow (TWV). 
Data are presented for both GO and CNC in terms of rotational frequency $f_{\varOmega}$ of the Taylor-Couette cell, 
Reynolds number $Re$, Taylor number $Ta$, and rotational supercritical Peclet number $Pe_{r,sc}$.}
\begin{tabular}{llllllll}
Sample &  Instability & $f_{\Omega}$ & $Re$ & $Ta$ & $Pe_{r,sc}^{(w)}$ \\
\midrule
GO & LCF  & 0.47~Hz  & 60 & 209 &  $-$  \\
  &  TVF  & 1.08~Hz  & 112  & 1237 & $-$\\
   &  WVF &  1.34~Hz  & 142  & 1954 &  20 \\
   &  MWV &  2.17~Hz  & 237 & 5465 & 41 \\
     & TWV  & 3.27~Hz  & 368  & 13194 & 64 \\
\hline
CNC  & LCF  & 3.25~Hz  & 124 & 1267 &  $-$  \\
    & TVF  & 6.18~Hz &  248 & 6009 &   $-$ \\
  & MWV &  7.05~Hz & 289 & 8214 &   0.096 \\
  & MWV &  8.65~Hz & 370 & 13332 &  0.10 \\
  & TWV  & 11.49~Hz & 516 & 25952 &  0.13 \\
\bottomrule
\end{tabular}
\label{TAB:instabilities}
\end{table}

%%%%%%%%%%%%%%%
%%% FIGURES %%%
%%%%%%%%%%%%%%%

% FIG 1: setup

% FIG 2: multiscale analysis

% FIG 3: static experiment

% FIG 4: multiscale data 

% FIG 5: frequencies

%SUPPLEMENTARY FIGs

%\setcounter{figure}{0}
\renewcommand{\figurename}{Supplementary Fig.}
\renewcommand{\thefigure}{S\arabic{figure}}

% FIG SI 6 - checks
\begin{figure}[t]
\centering
\includegraphics[width=\textwidth]{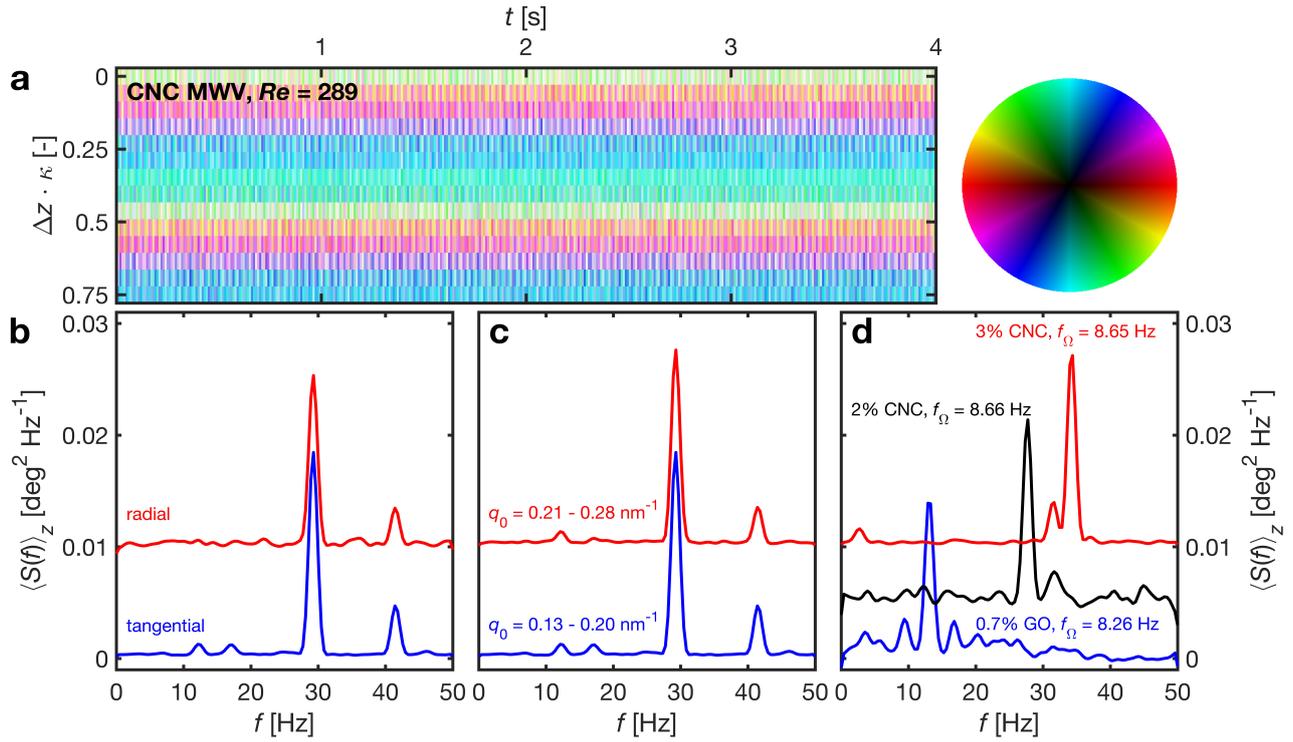}
\caption{Time-resolved scanning SAXS microscopy of nanoparticles under flow. 
\textbf{a}, Alternative FFT-based representation of the CNC MWV data in Fig.~\ref{FIG:SAXS}, with preferential 
alignment ($\varphi$), degree 
thereof ($\vert a_1\vert / \vert a_0 \vert$), and scattering power ($ \vert a_0 \vert$) encoded in the hue 
(cf. colour wheel), saturation, and value, respectively. The ensuing spatially averaged power spectral density $\langle S(f)_z\rangle$ 
is presented for (\textbf{b})~different scattering configurations and (\textbf{c})~different scattering-vector range $q_0$, 
exhibiting no effect of scattering configuration or scattering-vector modulus. 
\textbf{d} Comparison of $\langle S(f)_z\rangle$ for different suspensions - 0.7 wt\% GO, 3 wt\% CNC, and 2 wt\% CNC - 
at comparable rotational frequency $f_\Omega$ of the inner cylinder. The CNC suspensions show similar, fast nanoscale dynamics, 
whereas GO exhibits slower wavy dynamics, similar to what is expected based on the data of Fig.~\ref{FIG:freq}.  
The curves of panels~\textbf{b-d} have been rescaled and vertically offset for clarity. 
}
\label{FIG:checks}
\end{figure}

% FIG SI 7 - SAXS
\begin{figure}[t]
\centering
\includegraphics[width=0.7\textwidth]{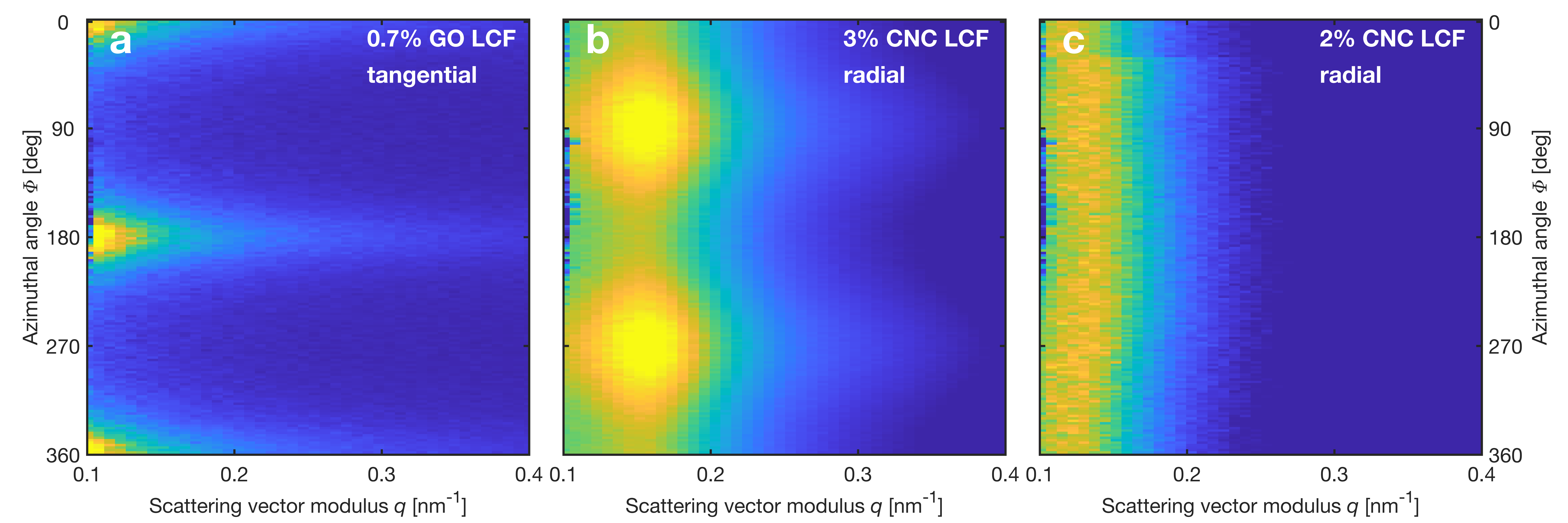}
\caption{Temporally averaged SAXS data acquired under laminar Couette flow (LCF). Data are 
collected from (\textbf{a}) GO in tangential and (\textbf{b}) CNC in radial scattering configuration, both 
shown as function of scattering vector modulus $q$ and azimuthal angle $\varPhi$. \textbf{c}, Data for a 
less concentrated CNC suspension are presented for comparison. 
}
\label{FIG:SAXS_LCF}
\end{figure}

% FIG SI 8 - viscosity
\begin{figure}[t]
\centering
\includegraphics[width=0.5\textwidth]{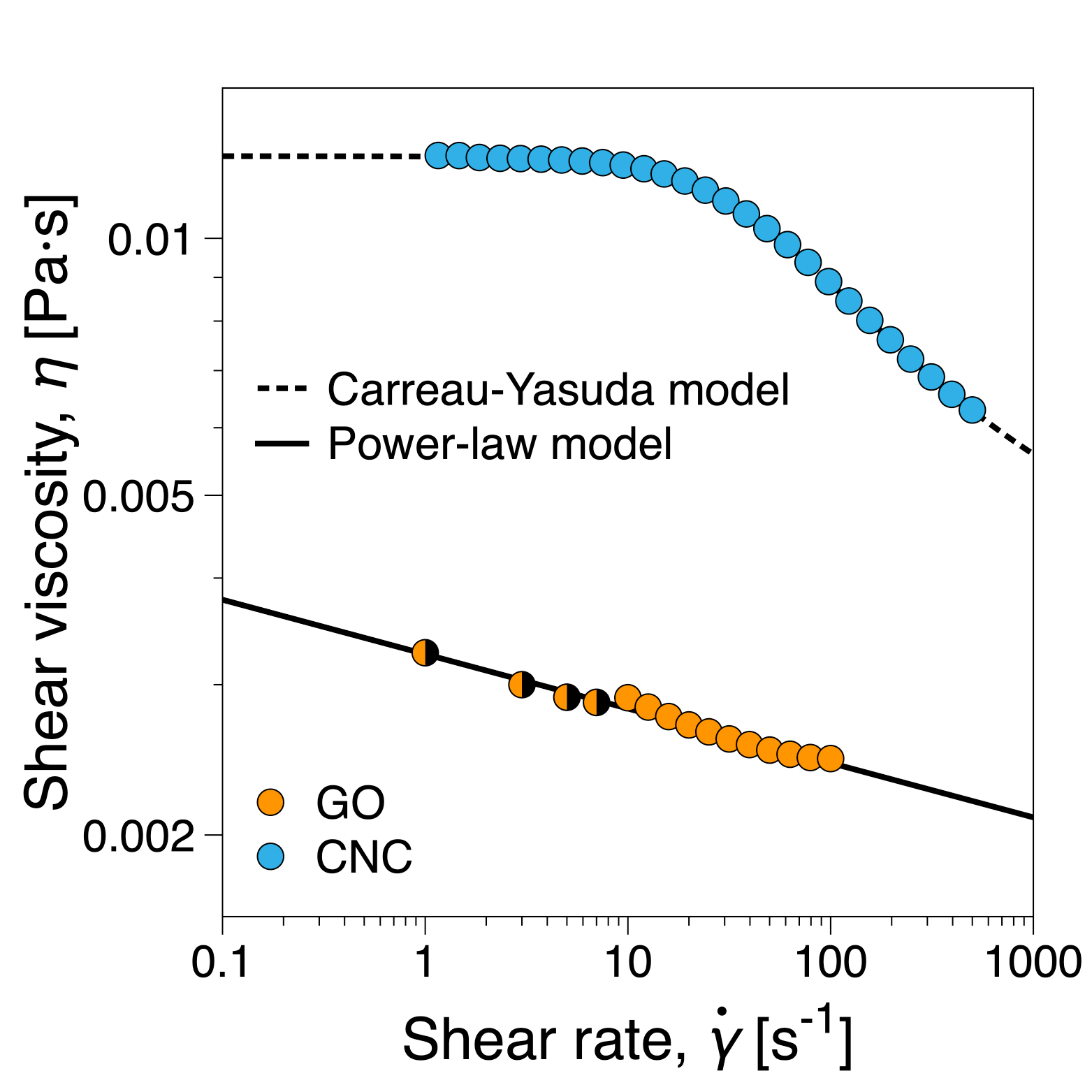}
\caption{Steady shear viscosity of the investigated suspensions. The half-filled symbols 
refers to long-term step input measurements, with each point starting for a fresh sample, in order to 
accurately reach steady-state. The 0.7 wt\% GO and 3 wt\% CNC data were fitted with the power-law and 
Carreau-Yasuda models, respectively.
}
\label{FIG:viscosity}
\end{figure}

% FIG SI 9 - spatiotemp
\begin{figure}[t]
\centering
\includegraphics[width=\textwidth]{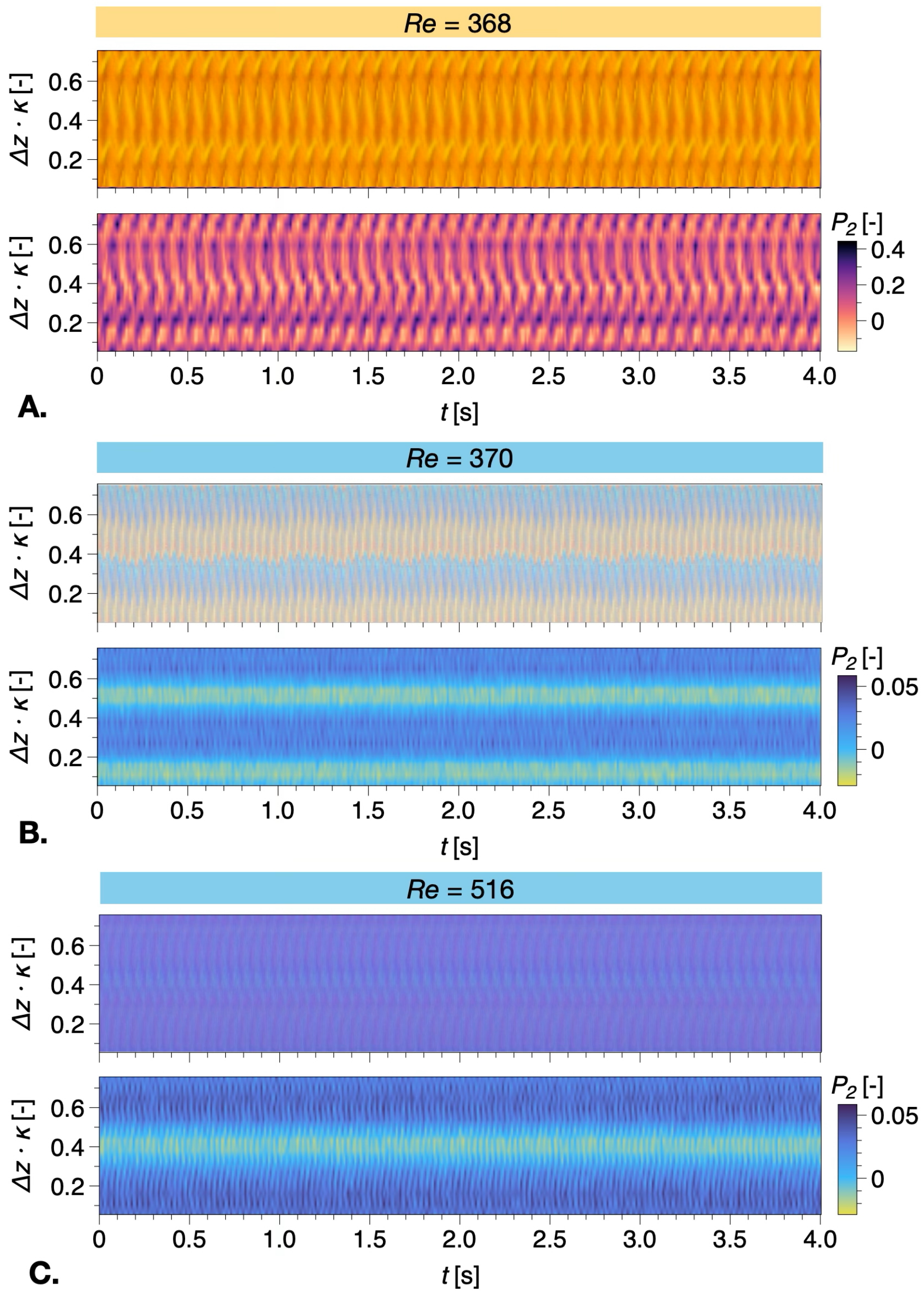}
\caption{Complementary data to the multiscale spatiotemporal analysis presented in Fig.~\ref{FIG:SAXS}. 
\textbf{a}, GO, \textbf{b-c} CNC.
}
\label{FIG:multiscale_complementary}
\end{figure}

\end{document}